\documentclass[aps,preprint,onecolumn]{revtex4-1}
\usepackage{graphicx}
\usepackage{subfig}
\usepackage{color}

\begin{document}

\title{Kinetics of random sequential adsorption of nearly spherically symmetric particles}

%
\author{Micha\l{} Cie\'sla$^{1}$}
 \email{michal.ciesla@uj.edu.pl}

\author{Jakub Barbasz$^{1,2}$}
\email{ncbarbas@cyf-kr.edu.pl}
\affiliation{
$^1$ M. Smoluchowski Institute of Physics, Jagiellonian University, 30-059 Kraków, Reymonta 4, Poland. \\
$^2$ Institute of Catalysis and Surface Chemistry, Polish Academy of Sciences, 30-239 Kraków, Niezapominajek 8, Poland.
}%
\date{\today}


\begin{abstract}
Kinetics of random sequential adsorption (RSA) of spheres on flat, two-dimensional surfaces is governed by a power law with exponent $-1/2$. The study has shown that for RSA of nearly spherically symmetric particles this exponent is $-1/3$, whereas other characteristics typically measured in RSA simulations approach values known for spheres with the increase of symmetry of the particles.
\end{abstract}

\maketitle
\section{Introduction}
Random sequential adsorption (RSA) is one of the simplest numerical algorithms used for modelling of irreversible adsorption processes. It is based on subsequent attempts to add a randomly placed particle to an adsorption layer. If the particle does not overlap with any of previously added particles it is added to the layer. Otherwise it is removed. One of the most important characteristics of an adsorption layer is its saturated coverage ratio
\begin{equation}
  \theta_{\rm max}  = n_{\rm max} \, \frac{S_{\rm M}}{S_{\rm C}},
\end{equation}
where $n_{\rm max}$ is number of adsorbed particles when any further adsorption act is not possible, $S_M$ is the surface covered by a single particle and $S_C$ is an area of underlying surface. Actually, to be sure that a coverage is saturated, the infinite number of algorithm steps is needed. To determine it from finite time simulations, kinetics of RSA layer growth has to be known. Since the very first studies by Feder \cite{bib:Feder1980} on spherical particles undergoing RSA procedure to form an adsorption layer on a two dimensional flat collector, it has been observed that the coverage ratio kinetics is governed by the following power law
\begin{equation}
\theta(t) = \theta_{\rm max} - A \cdot t^{-\frac{1}{d}},
\label{eq:fl}
\end{equation}
where $\theta(t)$ denotes the ratio of space covered by adsorbed particles to the whole space of a collector after $t$ algorithm steps, $\theta_{\rm max} \equiv \theta(t \to \infty)$ is saturated coverage ratio, $A$ is some constant and $d=2$. For spherically symmetric particles, the above relation was analytically confirmed valid also by other investigators \cite{bib:Swendsen1981,bib:Privman1991} and since then parameter $d$ is known to be equal to the dimension of a collector \cite{bib:Torquato2006,bib:Torquato2013} which can also be a fractal \cite{bib:Ciesla2012fractal,bib:Ciesla2013sponge}. The situation changes slightly for RSA of elongated particles e.g. spheroids, spherocylinders, rectangles, needles and similar \cite{bib:Talbot1989,bib:Vigil1989,bib:Tarjus1991,bib:Viot1992,bib:Ricci1992} and even for fibrinogens \cite{bib:Adamczyk2010}. In all these cases parameter $d$ in Eq.(\ref{eq:fl}) has been found to be equal to $3$ as long as particles are stiff \cite{bib:Ricci1992,bib:Ciesla2013polymer}. Interestingly, latest studies show that $d=3$ also for tetramers \cite{bib:Ciesla2013tetramers} as well as for hexamers \cite{bib:Ciesla2013trihex}. As some of these shapes are often approximated by spheres for numerical modelling purposes \cite{bib:Aznar2012,bib:Finch2013},  
it is possible that results obtained for spheres will, by default, be applied to the properties of such spherical-like particles \cite{bib:Ciesla2012dimers,bib:Ramsden1994}, which, could lead to wrong conclusions. Therefore, the aim of this study is to check which exponent $d$ describes RSA kinetics of nearly spherically symmetric molecules. To achieve this, a number of RSA simulation for particles of the growing number of symmetry axes was performed and analysed.
\section{Model}
\begin{figure}[htb]
\vspace{1cm}
\centerline{%
\includegraphics[width=0.8\columnwidth]{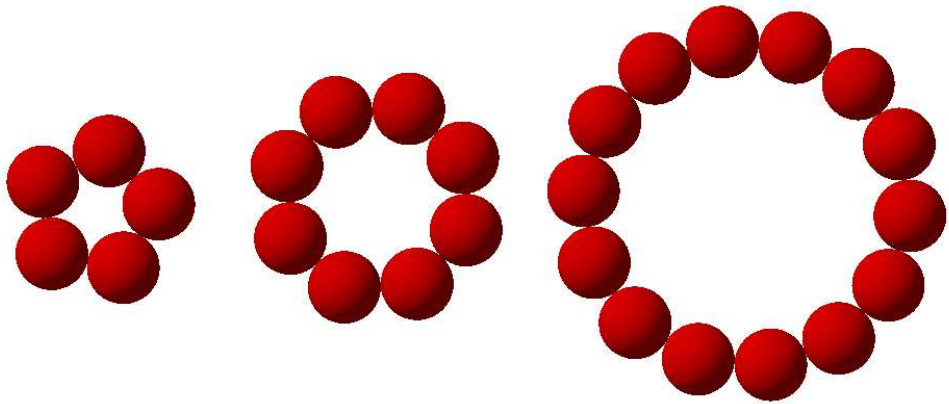}}
\caption{Example of rings built of 5, 8 and 14 identical spheres.}
\label{fig:model}
\end{figure}

Adsorbate particles are rings built of 5 to 40 identical spheres of radius $r_0$ was performed and analysed.  An example rings is presented in Fig.\ref{fig:model}.
Such particles were thrown randomly on a flat square collector of a side size of $1000\, r_0$ according to RSA procedure \cite{bib:Feder1980, bib:Ciesla2013tetramers, bib:Ciesla2013trihex}. Separate experiments were performed for particles of different sizes. During the simulation, temporary number of adsorbed particles $n(t)$ has been measured. For each type of molecules $100$ independent simulations were performed and each of them included $10^5\,t_0$ steps, where $t_0 = 10^6 / N \pi r_0^2$ is a dimensionless time unit equal to the ratio of collector surface to single particle surface. $N$ denotes the number of spheres in a ring. Note that any specific value of the time unit does not affect exponent $d$ in Eq.(\ref{eq:fl}) as long as it is proportional to the number of algorithm steps.
\section{Results}
\par
\begin{figure}[htb]
\vspace{1cm}
\centerline{%
\includegraphics[width=0.8\columnwidth]{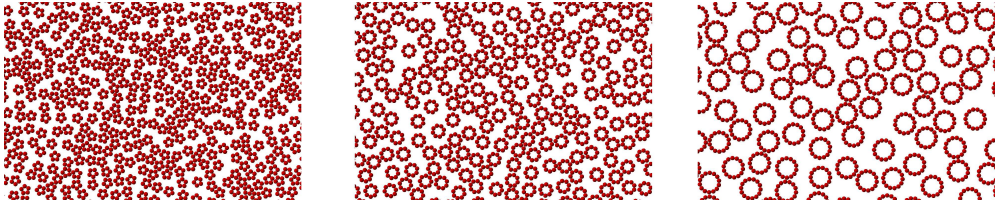}}
\caption{Examples of monolayers obtained using RSA procedure for rings built of 5, 8 and 14 identical spheres.}
\label{fig:layers}
\end{figure}

Examples of monolayers built of different size rings are presented in Fig.\ref{fig:layers}.
As the figure contains only a small fragment of the whole layer it is worth to mention that the average number of adsorbed rings on the whole collector was  $28857$, $14691$ and  $6020$ for rings built of $5$, $8$ and $14$ spheres, respectively. The standard deviation did not exceed 10 particles.
%
%
\par
\begin{figure}[htb]
\vspace{1cm}
\centerline{%
\includegraphics[width=0.8\columnwidth]{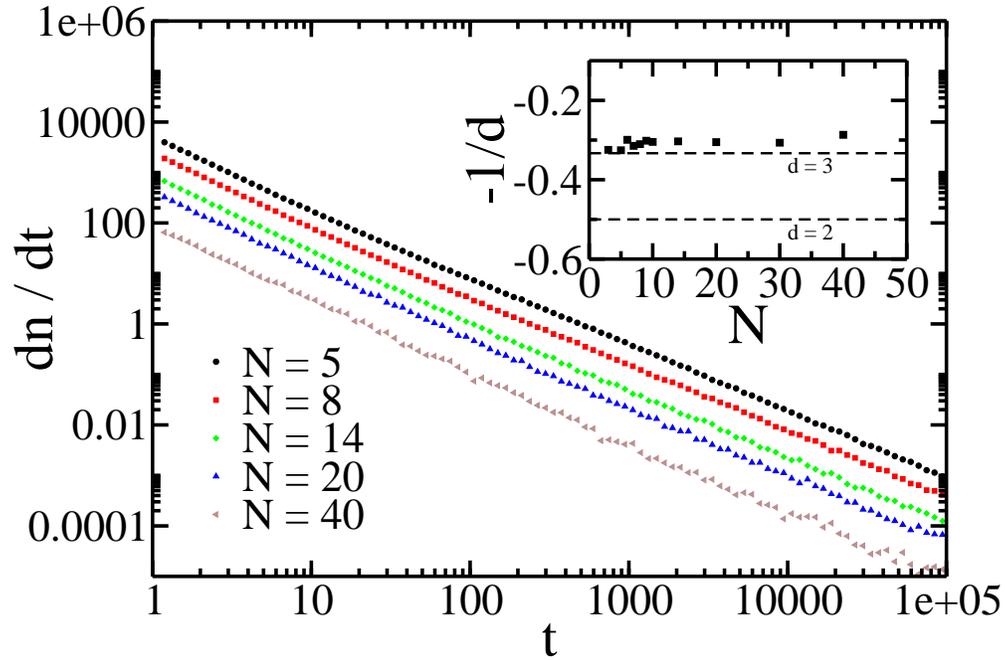}}
\caption{Increments of adsorbed particles versus number of RSA steps expressed in dimensionless time units $t_0$ for different sizes of rings. Inset shows the dependence of the exponent in Eq.(\ref{eq:fl}) on ring size. Statistical error is smaller than the size of squares. Horizontal dashed lines correspond to $d=2$ and $d=3$.}
\label{fig:kinetics}
\end{figure}

{\em RSA kinetics} measured for rings of different sizes are presented in Fig.\ref{fig:kinetics}.
Validity of Eq.(\ref{eq:fl}) has been confirmed for a wide range of simulation times. Exponent $d$, which corresponds to the slope of the lines in Fig.\ref{fig:kinetics}, almost does not depend on ring size and, within the studied range, is close to $d=3$, even for the largest and most spherically symmetric particles. This is highly unexpected result because with particle shape approaching sphere exponent $d$ should approach the value of $2$. However, it is generally possible that large rings do not approach spheres in terms of properties measured using RSA simulations. To check if this is the case, other characteristics typically obtained from RSA simulations were measured and compared with the ones obtained from the RSA of spheres.
\subsection{Saturated random coverage ratio.}
\par
\begin{figure}[htb]
\vspace{1cm}
\centerline{%
\includegraphics[width=0.8\columnwidth]{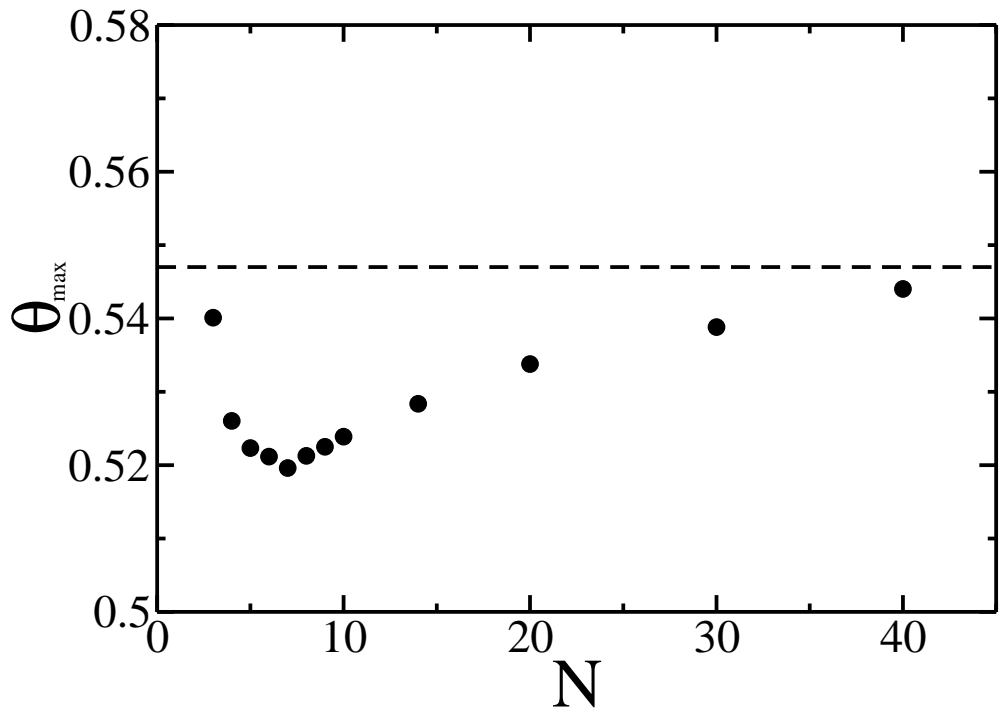}}
\caption{Saturated random coverage ratio for rings of different sizes. Dashed line corresponds to saturated random coverage ratio for a monolayer built of spheres \cite{bib:Torquato2006,bib:Ciesla2013polymer}. Statistical errors are smaller than the size of dots.}
\label{fig:qn}
\end{figure}

The surface covered by the ring is equal to $N\pi r_0^2$; however, the uncovered space inside the ring is also not available for subsequent particles adsorption. To compare obtained coverages with the spheres adsorption case, the interior of the ring should also be counted as covered. Therefore, the total collector area occupied by a single ring built of $N$ spheres is
\begin{equation}
S_{\rm M}(N) = N r_0^2\left[\cot\left( \frac{\pi}{N} \right) + \pi\frac{N+2}{2N} \right].
\end{equation}
As mentioned at the begining, the RSA simulation approaches saturated coverage after an infinite number of algorithm steps. Therefore to find $\theta_{\rm max}$ the Eq.(\ref{eq:fl}) is needed. Having determined the exponent $d$, let $y=t^{-1/d}$. Then Eq.(\ref{eq:fl}) converts into $\theta(y) = \theta_{\rm max} - A y$, where $A$ is a constant coefficient. Approximation of the linear relation for $y=0$ gives the saturated random coverage $\theta_{\rm max} \equiv \theta(y=0)$. Figure \ref{fig:qn} presents saturated random coverage ratios for different ring sizes.
Data for $N=3$ and $N=4$ were taken from \cite{bib:Ciesla2013trihex} and \cite{bib:Ciesla2013tetramers} respectively. For trimers ($N=3$), random saturated coverage ratio is only slightly lower than for spheres. Significant drop of $\theta_{\rm max}$ for larger N is probably connected with the peculiar shape of medium size rings, which effectively block slightly more space. For larger rings, saturated random coverage ratio, as expected, grows up to the value known for spheres.
\subsection{Available surface function.}
\par
\begin{figure}[htb]
\vspace{1.2cm}
\centerline{%
\includegraphics[width=0.8\columnwidth]{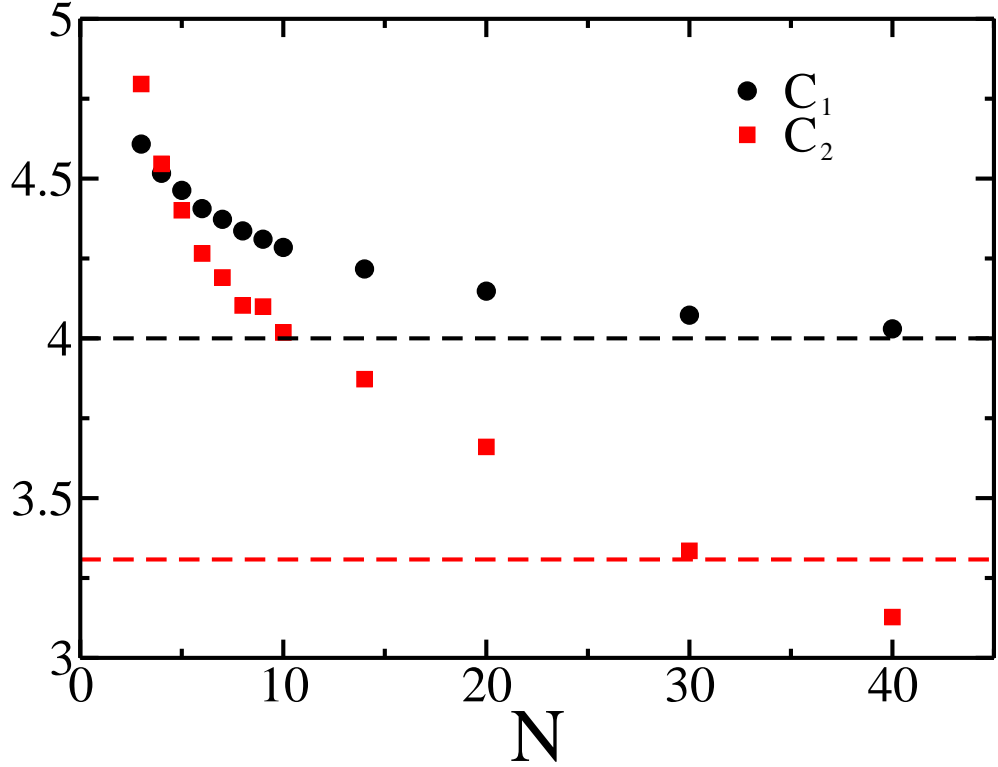}}
\caption{The $ASF(\theta)$ coefficients $C_1$ and $C_2$ for different ring sizes. Dashed lines correspond to their values for spheres: $C_1=4$ and $C_2 = \frac{6\sqrt{3}}{\pi}$. Statistical errors are smaller than the size of dots and squares.}
\label{fig:c1c2}
\end{figure}

Available surface function can be defined as a probability of finding an uncovered space large enough to place there a subsequent particle. For small coverages, it can be approximated as
\begin{equation}
{\rm ASF}(\theta) = 1 - C_1 \theta + C_2 \theta^2 + o(\theta^2).
\label{eq:asf}
\end{equation}
Expansion coefficient $C_1$ corresponds to the area blocked by a single particle, whereas $C_2$ corresponds to a cross-section of the surface blocked by two independent rings. Note that both of them are directly related to the second $B_2=1/2C_1$ and third $B_3=1/3C_1^2-2/3C_2$ viral coefficient of the equilibrium monolayer built of such particles \cite{bib:Tarjus1991,  bib:AdamczykBook}. 
Parameters $C_1$ and $C_2$ for rings were determined by fitting the Eq.(\ref{eq:asf}) to the simulation data. Results presented in Fig.\ref{fig:c1c2} show monotonic decrease of both the parameters down to the analitical values characterising RSA of spheres. In case of $C_2$, the parameter drops even significantly below the expected value; however, it should be noted that expansion of $ASF(\theta)$ up to the second order is valid only for small $\theta$ and estimation of $C_2$ is much more sensitive to that range than of $C_1$. Here for fitting purposes, we assumed that $\theta < 0.2 \, \theta_{\rm max}$.
\subsection{Density autocorrelation function}
\par
\begin{figure}[htb]
\vspace{1cm}
\centerline{%
\includegraphics[width=0.8\columnwidth]{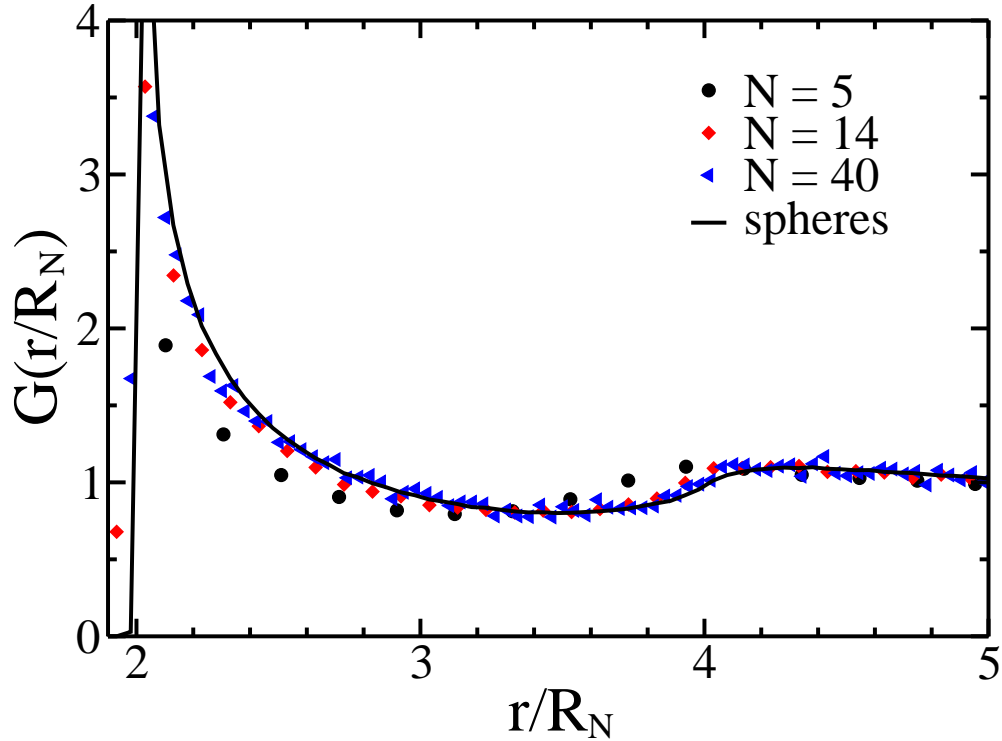}}
\caption{Density autocorrelation function for different ring sizes. The solid line corresponds to density autocorrelation function for spheres.}
\label{fig:corr}
\end{figure}

Density autocorrelation function gives an insight into coverage structure and is defined as:
\begin{equation}
G(r) = \frac{P(r)}{2\pi r \rho},
\end{equation}
where $P(r)dr$ is a probability of finding two particles in a distance between $r$ and $r+dr$. Here, the distance $r$ is measured between the geometric centres of molecules.  As $\rho$ is the mean density of particles inside a covering layer, thus $G(r\to \infty) = 1$. To compare density autocorrelations for different ring sizes, the length has to be rescaled and hence $r \to r/R_N$ where
\begin{equation}
R_N = r_0 \left( \frac{1}{\sin\frac{\pi}{N}} + 1 \right)
\end{equation}
is radius of the ring built of $N$ spheres. The comparison of $G(r/R_N)$ is presented in Fig.\ref{fig:corr}.
Again, values obtained for the largest rings approach the limit given by $G(r)$ for spheres. This is yet another indication that random coverage properties for large rings match the ones for spheres.
\section{Discussion}
\par
\begin{figure}[htb]
\vspace{1cm}
\centerline{%
\includegraphics[width=0.8\columnwidth]{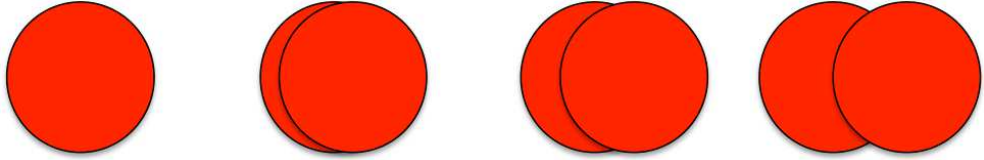}}
\caption{Examples of particles built of two, partially overlapped spheres for anisotropy parameter $\epsilon$ equal to (from left) $0$, $0.1$, $0.25$ and $0.5$, respectively.}
\label{fig:odimer}
\end{figure}
\begin{figure}[htb]
\vspace{1cm}
\centerline{%
\includegraphics[width=0.8\columnwidth]{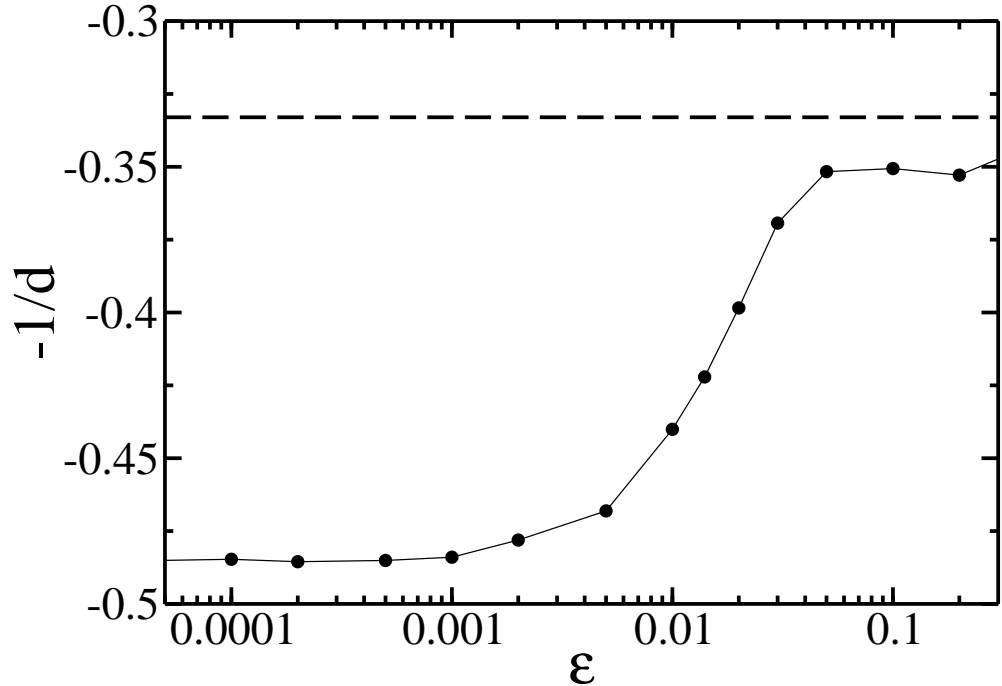}}
\caption{The exponent from Eq.(\ref{eq:fl}) dependence on anisotropy parameter $\epsilon$. Dashed line corresponds to $d=3$. The statistical errors of presented data are below $0.003$ and they are smaller than size of dots. The full connecting dots has been drawn to guide the eye.}
\label{fig:odimerfl}
\end{figure}

The difference between RSA kinetics for spheres and for nearly spherically symmetric particles is counterintuitive, especially when considering results obtained by Viot et al. \cite{bib:Viot1992} for convex particles. In that study, RSA kinetics for particles having the length to height ratio below $1.25$ was closer to one for spheres ($d=2$) than for elongated particles $d=3$. However, the behaviour of RSA for concave particles can be significantly different \cite{bib:Jao2008, bib:Schelke2008}. Therefore, to find out where in this case the transition from $d=2$ to $d=3$ should occur in our case, we studied RSA for particles built of two identical, partially overlapped spheres (see Fig.\ref{fig:odimer}). As the ratio of particle width to height is $(1+\epsilon)$ the parameter $\epsilon$ can be used as an anisotropy measure.
The RSA kinetics exponent defined in Eq.(\ref{eq:fl}) for different $\epsilon$ is shown in Fig.\ref{fig:odimerfl}.
The transition from $d=3$ to $d=2$ for partially overlapped spheres begins at $\epsilon \approx 0.02$ which is an order of magnitude lower than in case of convex ellipsoids or spherocylinders \cite{bib:Viot1992}. It is worth to notice that even for $\epsilon=0.1$ when particle looks as almost spherical (see. Fig.\ref{fig:odimer}) the RSA kinetics still behaves as for elongated particles. In the case of previously simulated rings, the anisotropy reaches $0.02$ for $N\approx 160$, which partially explains why the the transition has not been observed in Fig.\ref{fig:kinetics}.   
\section{Summary}
\par
Although properties of saturated random coverages built of spheres and nearly spherically symmetric particles are almost the same, the RSA kinetics is significantly different. Therefore, to obtain saturated random coverage ratio $\theta_{\rm max}$ on a flat two-dimensional surface from a finite time simulation, the $-1/3$ exponent in Eq.(\ref{eq:fl}) should be used instead of $-1/2$, which characterises RSA of spheres. Although the difference between this two approximations is currently at the border of accuracy of typical experiments, development of new experimental techniques could make this difference significant. It was also shown that other characteristics typically measured in RSA simulation approach the value for spheres along with the growth of particle symmetry level.
\section*{Acknowledgements}
This work was supported by Polish National Science Centre grant no. UMO-2012/07/B/ST4/00559.
%


\begin{thebibliography}{20}
\bibitem{bib:Feder1980} J. Feder, J. Theor. Biol. {\bf 87}, 237 (1980).
\bibitem{bib:Swendsen1981} R. H. Swendsen, Phys. Rev. A {\bf 24}, 504 (1981).
\bibitem{bib:Privman1991} V. Privman, J.-S. Wang, and P. Nielaba, Phys. Rev. B {\bf 43} 3366 (1991).
\bibitem{bib:Torquato2006} S. Torquato, O.U. Uche, F.H. Stillinger, Phys.Rev.E {\bf 74} 061308 (2006).
\bibitem{bib:Torquato2013} G. Zhang, S. Torquato, Phys. Rev. E {\bf 88}, 053312 (2013).
\bibitem{bib:Ciesla2012fractal} M. Cie\'sla, J. Barbasz, J. Chem. Phys. {\bf 137} 044706 (2012).
\bibitem{bib:Ciesla2013sponge} M. Cie\'sla, J. Barbasz, J. Chem. Phys. {\bf 138} 214704 (2013).
\bibitem{bib:Talbot1989} J. Talbot, G. Tarjus, P. Schaaf, Phys. Rev. A {\bf 40} 4808 (1989).
\bibitem{bib:Vigil1989} R.D. Vigil R.M. Ziff, J. Chem. Phys. {\bf 91} 2599 (1989).
\bibitem{bib:Tarjus1991} G. Tarjus, P. Viot, Phys. Rev. Lett. {\bf 67} 1875 (1991)
\bibitem{bib:Viot1992} P. Viot, G. Tarjus, S. M. Ricci, J. Talbot, J. Chem. Phys. {\bf 97}, 5212 (1992).
\bibitem{bib:Ricci1992} S. M. Ricci, J. Talbot, G. Tarjus, P. Viot, J. Chem. Phys. {\bf 97}, 5219 (1992).
\bibitem{bib:Adamczyk2010} Z. Adamczyk, J. Barbasz, M. Cie\'sla, Langmuir {\bf 26} 11934 (2010).
\bibitem{bib:Ciesla2013polymer} M. Cie\'sla, Phys. Rev. E {\bf 87}, 052401 (2013).
\bibitem{bib:Ciesla2013tetramers} M. Cie\'sla, J. Stat. Mech. P07011 (2013).
\bibitem{bib:Ciesla2013trihex} M. Cie\'sla, J. Barbasz, J. Mol. Model. DOI: 10.1007/s00894-013-2031-5 (2013).
\bibitem{bib:Aznar2012} M. Aznar, A. Luque, D, Reguera, Phys. Biol. {\bf 9} 036003 (2012).
\bibitem{bib:Finch2013} C. Finch, T. Clarke, J.J. Hickman, J. Comput. Phys. {\bf 244} 212 (2013).
\bibitem{bib:Ciesla2012dimers} M. Ciesla, J. Barbasz, J. Stat. Mech., 03015 (2012).
\bibitem{bib:Ramsden1994} J.J. Ramsden, G.I. Bachmanova, A.I. Archakov, Phys. Rev. E {\bf 50} 5072 (1994).
\bibitem{bib:AdamczykBook} Z. Adamczyk, {\em Particles at Interfaces: Interactions, Deposition, Structure}, Elsevier/Academic Press, Amsterdam, 2006.
\bibitem{bib:Jao2008} Y. Jiao, F.H. Stillinger, S. Torquato, Phys. Rev. Lett. {\bf 100} 245504 (2008).
\bibitem{bib:Schelke2008} P.B. Shelke, S.B. Ogale, M.D. Khandkar, A.V. Limaye, Phys. Rev. E {\bf 77} 066111 (2008).
\end{thebibliography}
\end{document}